\documentclass[intlimits,twoside,a4paper]{article}

\usepackage[cp1251]{inputenc}

\usepackage[eqsecnum]{cmpj3}

\usepackage{bm}

\issue{2029}{29}{1}{13502}
\doinumber{10.5488/CMP.29.13502}

\title[Monohydric alcohols under pressure]
{Pressure effects 
in the properties of simple monohydric alcohols.
Lessons from molecular dynamics simulations of united atom type UAM-EW model
}

\author[M. Aguilar, L. Pusztai, O. Pizio]
{M. Aguilar\orcid{0000-0003-3850-1188}\refaddr{label1},
L. Pusztai\orcid{0000-0002-0560-9902}\refaddr{label2,label3},
O. Pizio\orcid{0000-0001-8333-4652}\refaddr{label1}
\thanks{Corresponding author: \email{oapizio@gmail.com}.}}

\addresses{
\addr{label1}
Instituto de de Qu\'{i}mica, Universidad Nacional Aut\'{o}noma de M\'{e}xico,
Circuito Exterior, 04510, Cd. Mx., M\'{e}xico
\addr{label2} Wigner Research Centre for Physics, H-1121, Budapest,
Konkoly Thege M. \'ut. 29-33, Hungary
\addr{label3} Faculty for Advanced Science and Technology, Kumamoto University,
2-39-1 Kurokami, Chuo-Ku, Kumammoto, 860-8555, Japan
}


\date{Received 5 November 2025; accepted 2 December 2025; published 30 March 2026}
\begin{document}

\maketitle

\begin{abstract}
We explore the pressure dependence of a set of
properties of simple monohydric alcohols, namely of methanol, ethanol and 
1-propanol,
by using isobaric-isothermal molecular dynamics computer simulations.
A recently proposed united atom, non-polarizable force field for each of alcohols
[\,V. Garc\'{i}a-Melgarejo et al., J. Mol. Liq., \textbf{323}, 114576 (2021)] is applied.
Accuracy of the force field is evaluated
by comparing the simulation results and available experimental data from
the literature. Specifically, the density of alcohols upon
increasing pressure, the isothermal compressibility, the
static dielectric constant and self-diffusion coefficient are investigated
starting from 1 bar up to 3 kbar.
Evolution of the microscopic structure under pressure is discussed
in terms of the pair distribution functions and some coordination numbers. 
Conclusions of the present modelling and necessary developments 
to consider in future work are commented on.

\printkeywords
%
%

\end{abstract}

\section{Introduction}

This manuscript is dedicated to the memory of our close friend
and long-term co-worker over the past decades
Prof. Stefan Soko\l owski who passed away in 2024, unfortunately.
We are deeply sad because of this loss.
Stefan Sokolowski made important contributions to the 
statistical mechanical theory
of fluids and mixtures. Specifically, his scientific interests
were focused on the theory of inhomogeneous associating fluids
and mixtures with and without electrostatic interactions
\cite{stefan1,stefan2,stefan3}.
He was eager to discover novel features in the behavior
of these systems and to develop appropriate 
methodological tools to describe them. Computer simulation
approaches, such as the Monte Carlo technique, molecular dynamics
and dissipative molecular dynamics, besides entirely theoretical methods, 
were the focus of his interests as well. His area of research also
involved analyses of relations between theories and experimental
studies~\cite{stefan4} and apparently very distant
complex systems~\cite{stefan5,stefan6,stefan7}.

Our present report is  motivated by several factors. 
Namely, in \cite{pusztai1} we explored the influence of pressure on some properties
of neat methanol very recently. Molecular dynamics simulations
have been performed by using three models of methanol to validate 
theoretical predictions for the microscopic structure in comparison with
the results of X-ray and neutron diffraction experiments from ambient 
pressure, 1 bar, up to gigapascal pressures.
This research has been stimulated by previous 
experimental and computer simulation investigations of the behavior
of water and water-alcohol mixtures at high pressures with particular
emphasis on the evolution of the hydrogen bonding network. 

On the other hand, previous work from this laboratory was focused
on compositional trends of the behavior and mixing of species 
in water-methanol mixtures dependent on temperature and pressure~\cite{vmt}. However,
several issues concerning pressure effects remained out of attention of the authors.

Finally, it is worth mentioning that
methanol-water liquid mixtures at room temperature and atmospheric pressure are
among the most extensively studied hydrogen-bonded liquids. 
In order to extend previous observations, composition aspects of mixing of water 
and alcohol species were investigated by considering ethanol (EtOH) and 
1-propanol (PrOH) rather than MeOH-water solutions in \cite{bermudez}.
That work, however, was restricted to ambient pressure, 1~bar, and room temperature, 298.15~K.

Having all the above mentioned issues in mind, in the present work, we would like
to present very fresh results concerning pressure effects on some 
basic properties of three alcohols, MeOH, EtOH and PrOH. We employ molecular
dynamics simulation as methodological tools and perform comparisons of
theoretical predictions with available experimental data from the literature.
We are convinced that insights into the microscopic structure, thermodynamic, 
dynamic and dielectric properties of the systems in question from simulations
not only complement  various experimental data
but lead to a more profound understanding of various properties.
Moreover, present simulation findings may stimulate experiments for the structure
at high pressures that are not available up to our best knowledge,
e.g. for neat EtOH and PrOH and their mixtures with water and/or
with organic liquids. In addition, the simulation results for a given model 
can definitely guide the development of a more sophisticated and more accurate force field
for alcohol species.

The most important, initial step of computer simulations methodology lies in
the design of an appropriate force field. The intramolecular structure of alcohol species
in many cases may be and is considered at 
different levels of sophistication, from simple united atom models to 
all-atom and  to polarizable ones. Ab initio calculations have been attempted for alcohols
as well.  However, an increasing level of complexity of the models is not necessarily accompanied by
the improvement of predictions for the basic properties.
Besides, sophisticated models require expensive calculations. 

In this work, we use a simple united atom model for each of alcohols under study,
namely the UAM-EW type model~\cite{melgarejo} (note that the potential model
denoted as UAM-I in \cite{pusztai1} is identical to UAM-EW). 
Its parametrization
at 298.15~K and ambient pressure, 1 bar, described 
very much in detail in~\cite{melgarejo} involves
the fluid density, $\rho$, the dielectric constant, $\varepsilon$, and the surface 
tension, $\gamma$, as target properties. 
In addition, the model for each alcohol
is adjusted to provide a correct miscibility with TIP4P-$\varepsilon$
water force field~\cite{ale1}. One of the essential merits of the UAM-EW model,
in contrast to other models of this type, is that it
reproduces three target properties simultaneously, with reasonable accuracy
(deviations from the experimental data are less than a few percent) and correct
miscibility with water is ensured. However, a detailed analysis of the
performance of the model dependent on temperature and pressure has not been
performed so far.

\section{Models and simulation details}

The united atom type, non-polarizable, UAM-EW model~\cite{melgarejo}
for alcohols assumes sites, O, H, CH$_2$ and CH$_3$, and
is constructed from the interaction  potential between all atoms and/or groups. It is considered
as a sum of Lennard-Jones (LJ) and Coulomb terms.
All the parameters for inter- and intra-molecular interactions are given in the
supplementary material to~\cite{melgarejo}.
Lorentz-Berthelot combination rules were used to determine cross parameters for
the relevant potential well depths and diameters.

Molecular dynamics computer simulations were performed in the
isothermal-isobaric (NPT) ensemble at temperature $298.15$~K and at pressures in the interval 
between 1 bar and 3000 bar.
We used the GROMACS software package~\cite{gromacs} version 5.1.2.
The simulation box in each run was cubic, the total number of molecules
in all cases was fixed at 3000. 
As common, periodic boundary conditions were used.
Temperature and pressure control was provided by the V-rescale thermostat and Parrinello-Rahman
barostat with $\tau_T = 0.5$ ps and $\tau_P = 2.0$ ps, the timestep was 0.002 ps.

The non-bonded interactions were cut-off at 1.1 nm, whereas the long-range electrostatic interactions
were handled by the particle mesh Ewald method implemented in the GROMACS software package  (fourth
order, Fourier spacing equal to 0.12) with the precision $10^{-5}$. 
The van der Waals correction terms to the energy and pressure were applied.
In order to maintain the geometry of molecules the LINCS algorithm was used.

After preprocessing and equilibration, consecutive simulation runs with
the starting configuration being the last configuration from the previous
run, were performed to obtain trajectories for the data analysis.
The results for the majority of  properties  were obtained by averaging
over a set of runs. However, the dielectric constant was obtained from the
entire long trajectory. The self-diffusion coefficients of species were calculated from the
best slope of the mean squared displacement as common.

\section{Results and discussion}

\subsection{Density and isothermal compressibility}

\begin{figure}[h]
\begin{center}
\includegraphics[width=5.3cm,clip]{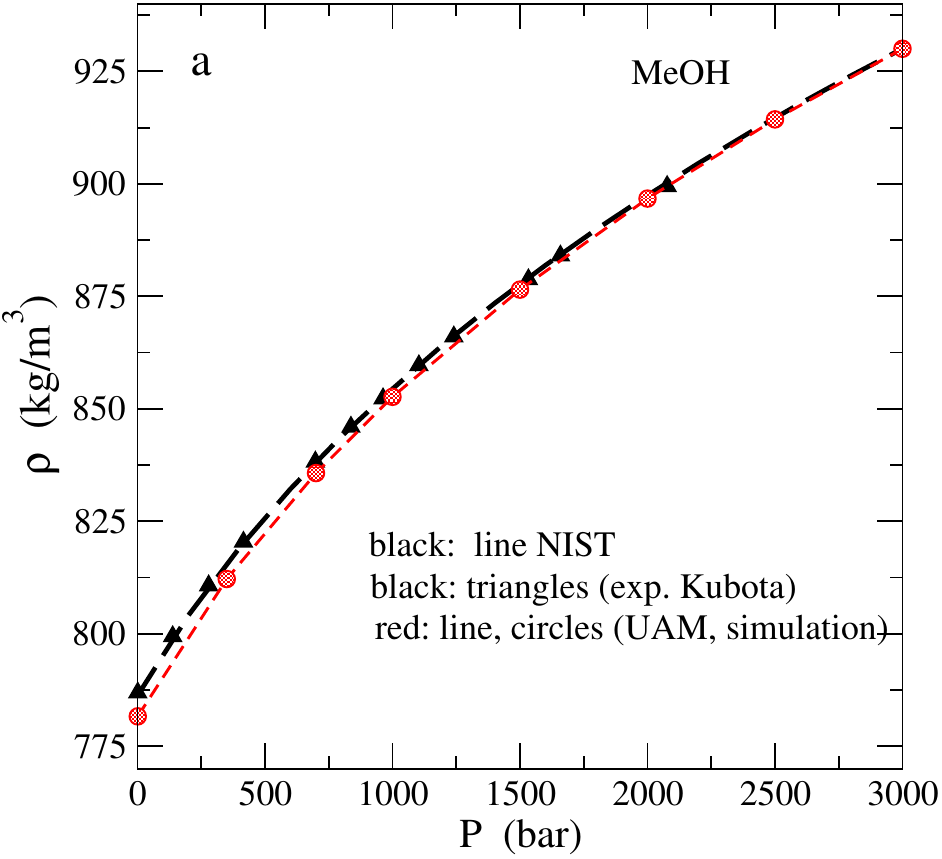}
\includegraphics[width=5.3cm,clip]{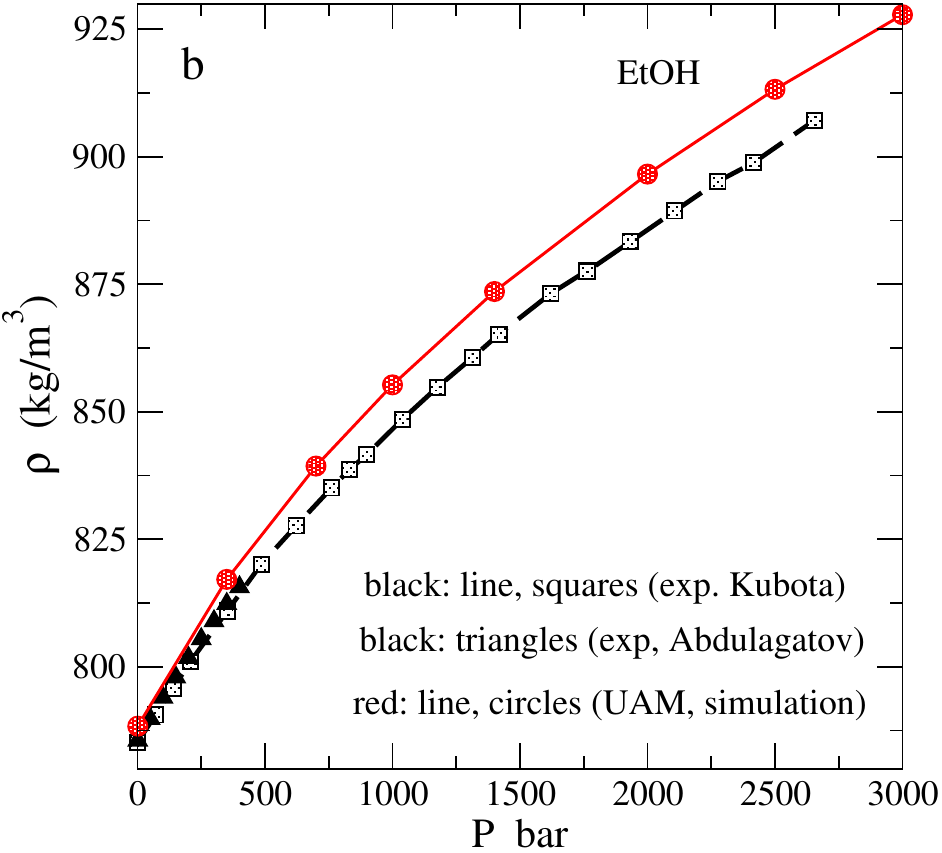}
\includegraphics[width=5.3cm,clip]{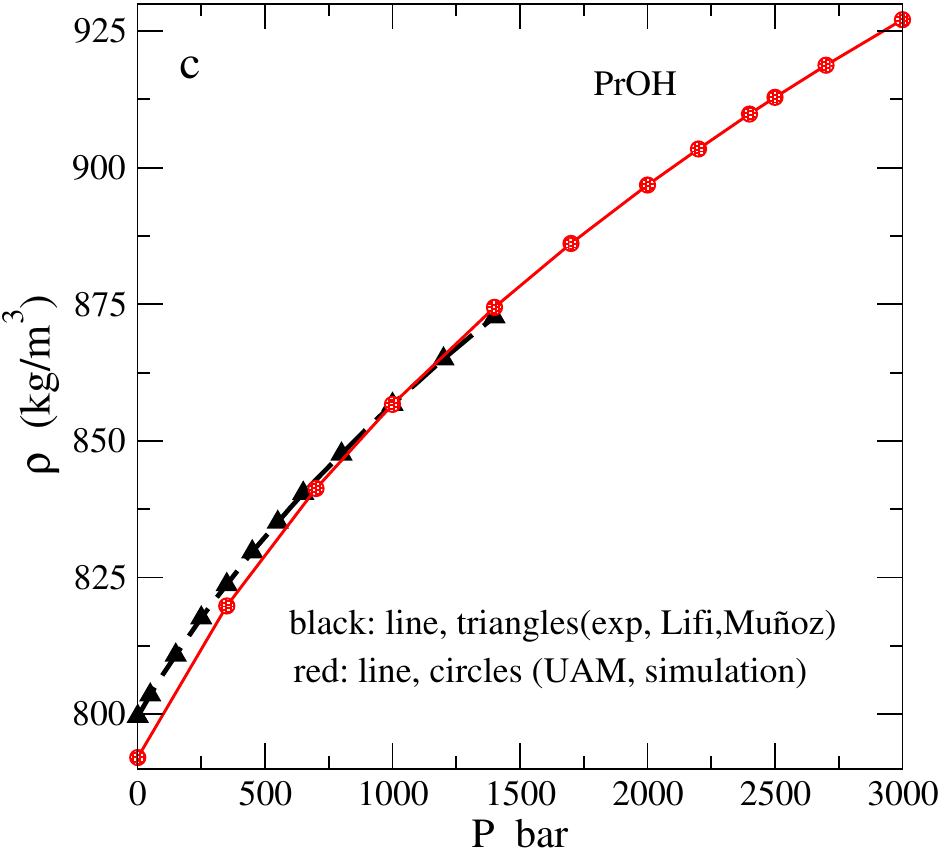}
\end{center}
\caption{(Colour online) Panels a, b and c: Methanol, ethanol and
propanol density on pressure at 298.15~K, respectively.
The simulation results are for the UAM-EW united atom models.
The experimental data are from
\cite{kubota-met,kubota-met2,nist} (panel a),
\cite{kubota-eth,abdul} (panel b), and \cite{munoz,lifi} (panel c).
}
\label{fig1}
\end{figure}

The dependence of density for each alcohol on
pressure from molecular dynamics simulations and comparison with experimental data
are shown in three panels of figure~\ref{fig1}. 

The plots for each alcohol are
shown separately for better visualization, because the 
absolute values are close to each other.
The experimental data in panel a are from \cite{kubota-met,kubota-met2,nist} and
in panel b from \cite{kubota-eth,abdul}. Finally, the experimental values of PrOH density on 
pressure are taken from \cite{munoz,lifi}. The best agreement between the simulation 
results and experimental points is observed for liquid MeOH. On the other hand, the
UAM-EW model overestimates the growth of the EtOH density with increasing pressure, in
comparison with experiments.
By contrast, the density for liquid PrOH is underestimated at low pressures. Better
agreement with experiment is observed at intermediate pressures, in the interval 
between 0.8 kbar and 1.4 kbar. Apparently, at high pressures the model
behaves reasonably, experimental data are not available for even higher pressures. 
Still, an overall performance of the UAM-EW  model can be termed as entirely
satisfactory for three alcohols over a wide pressure range.

\begin{figure}[h]
\begin{center}
\includegraphics[width=6.0cm,clip]{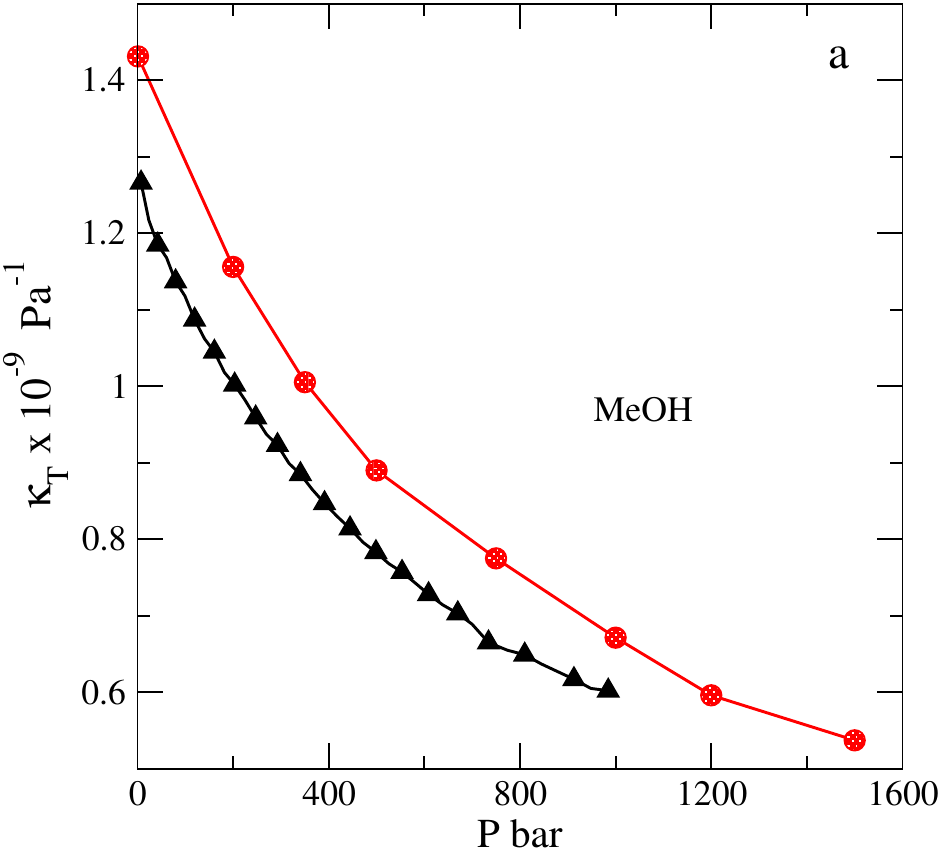}
\includegraphics[width=6.0cm,clip]{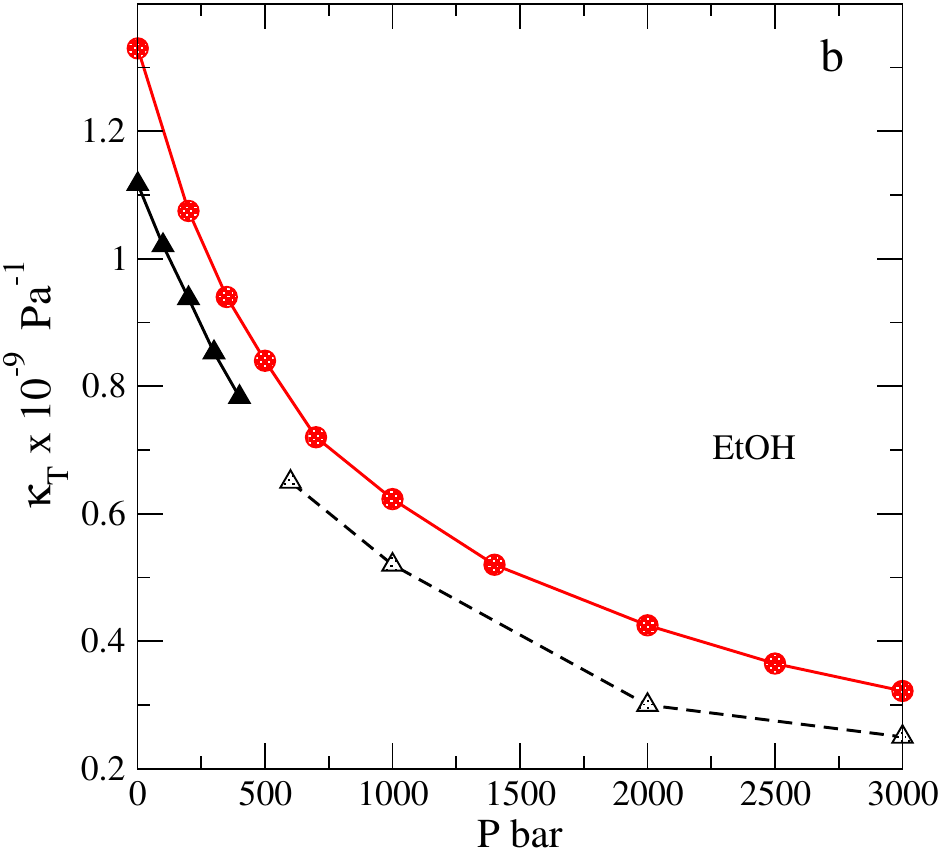}
\includegraphics[width=6.0cm,clip]{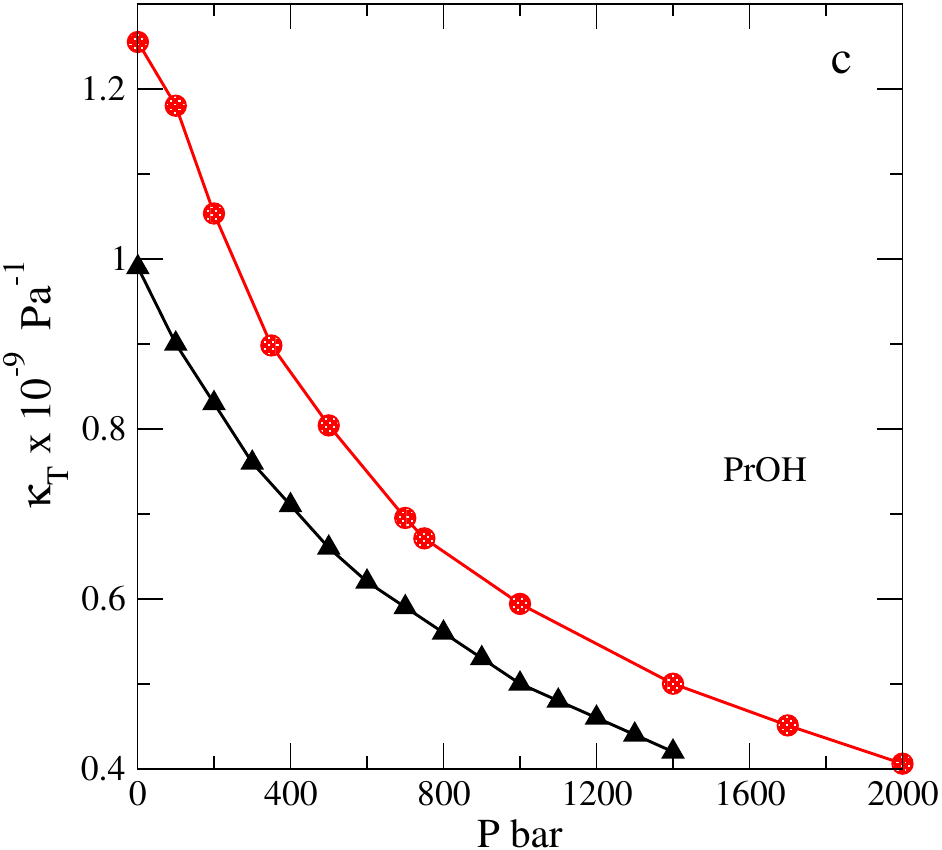}
\end{center}
\caption{(Colour online) Pressure dependence of the isothermal
compressibility for MeOH (panel a), EtOH (panel b), and PrOH (panel c).
The experimental data are from \cite{taravillo} for MeOH; from
\cite{pecar} (solid triangles)
and \cite{kubota-eth} (hollow triangles) for EtOH, respectively;
and from \cite{lifi} for PrOH.
}
\label{fig2}
\end{figure}

It is worth noting that in the case of MeOH, the TraPPE model \cite{trappe}  and the
MeOH OPLS/2016 model from \cite{salgado}, provide quite accurate
descriptions of density on pressure as documented in  our recent
contribution~\cite{pusztai1}. On the other hand, we are not aware of computer
simulation results for density on pressure using the TraPPE model for EtOH and PrOH.

In order to get additional insights into the performance
of the UAM-EW models for density in the temperature-pressure plane, 
we illustrate  the isothermal compressibility on pressure for MeOH, EtOH and PrOH in
figure~\ref{fig2}. From the previous figure~\ref{fig1}, that describes the $\rho(P)$ behavior, we learned that the 
discrepancy between simulation data and
experimental results is quite small for all thermodynamic states
studied. However, in all cases shown in figure~\ref{fig2}, the
differences between theory and experiment for isothermal compressibilities are more 
pronounced in terms of absolute values. Still, the shape of the curves for $\kappa_T (P)$ 
is very similar from simulations and experiment,
indicating an entirely satisfactory performance of the models for this specific property.
It is worth mentioning that isothermal $\rho(P)$ measurements 
fitted to various equations of state are frequently used
to elaborate the compressibility values in experimental works.  

We do not discuss trends of behavior of 
the isoentropic compressibility, $\kappa_S$, and the coefficient of 
isobaric thermal expansion, $\alpha_P$,
in the present work. Their behavior will be considered in a separate study
to explore temperature and pressure trends of a more complete set 
of properties related to fluctuations.
Now, we proceed to the results for the dielectric constant on pressure.
\newpage
\subsection{Dielectric constant}

The dielectric constant is one of the most important physico-chemical
properties of a given liquid, as it determines to a great extent the mixing
properties with other substances. 
Our results for $\varepsilon$  follow 
from the time-average of the fluctuations of the total
dipole moment of the system~\cite{martin}, as common in simulations,

\begin{equation}
\varepsilon=1+\frac{4\piup}{3k_{\text B}TV}\left( \left\langle\bf M^2\right\rangle-\big\langle\bf M\big\rangle^2\right) ,
\end{equation}
where $k_{\text B}$ is the Boltzmann constant and $V$ is the simulation cell volume.

\begin{figure}[h]
\begin{center}
\includegraphics[width=5.5cm,clip]{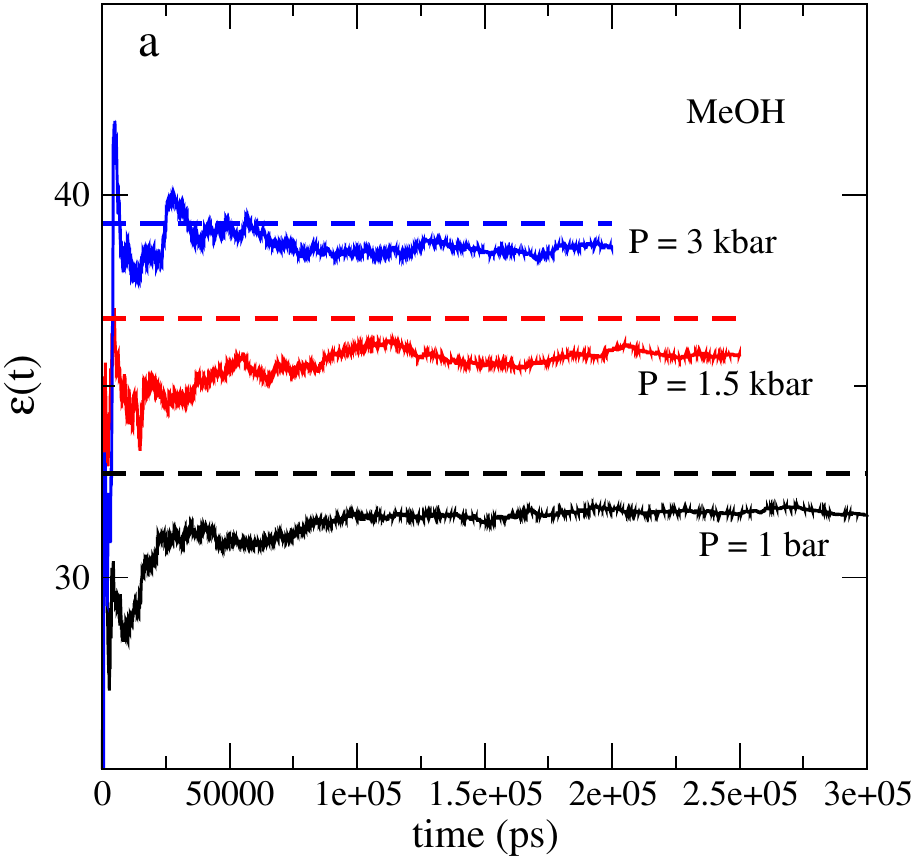}
\includegraphics[width=5cm,clip]{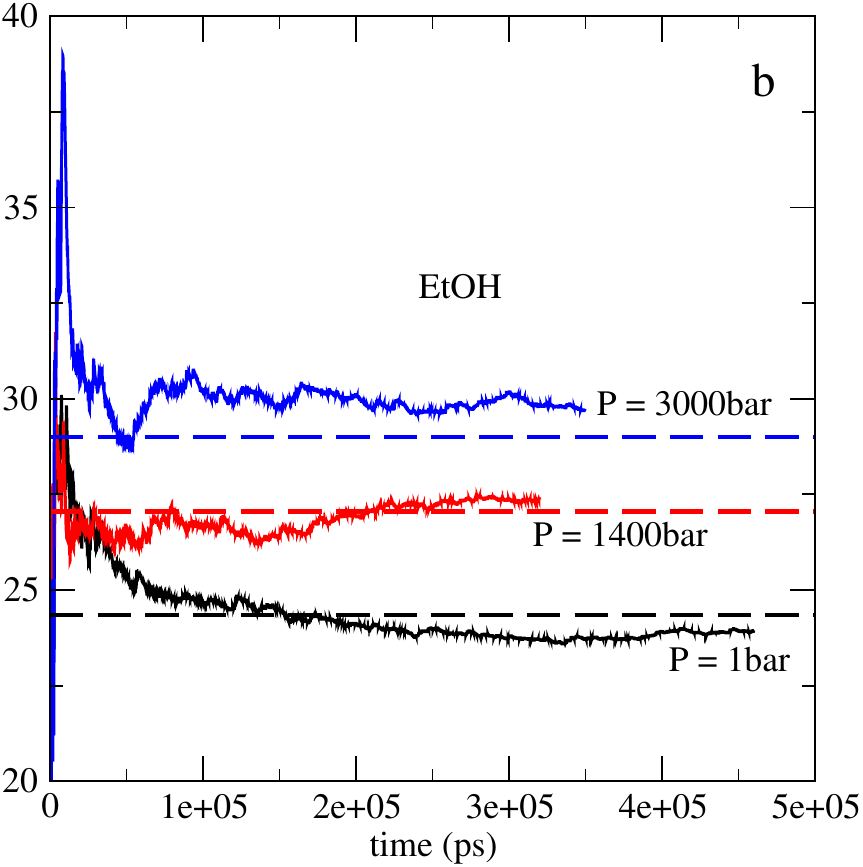}
\includegraphics[width=5cm,clip]{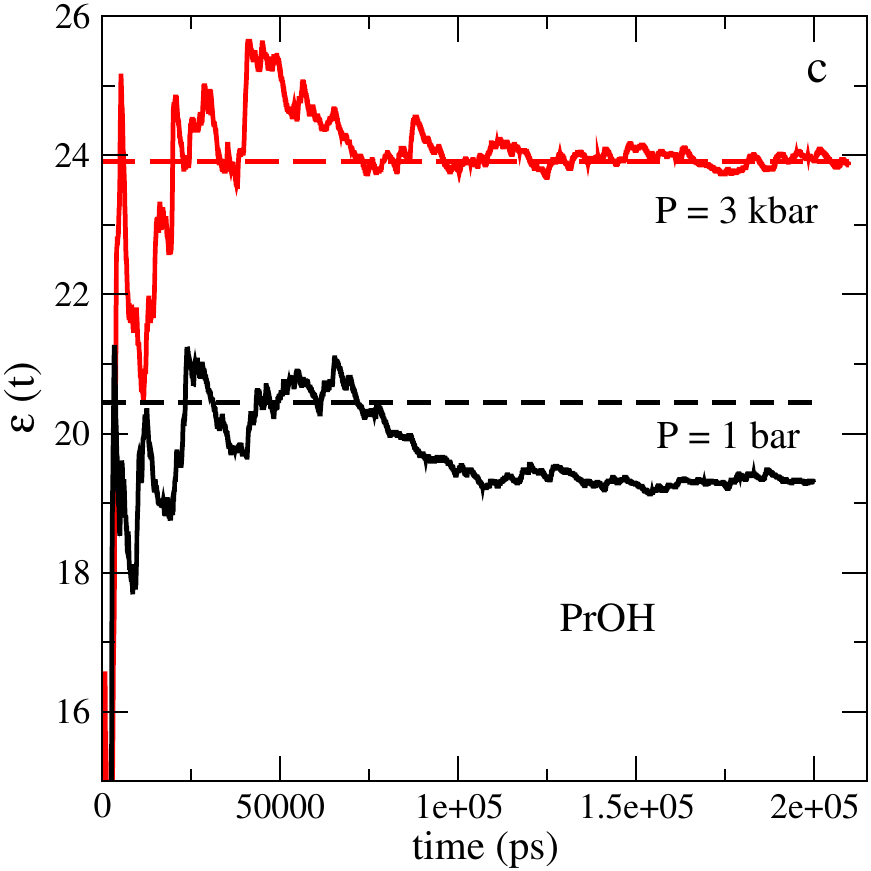}
\includegraphics[width=5cm,clip]{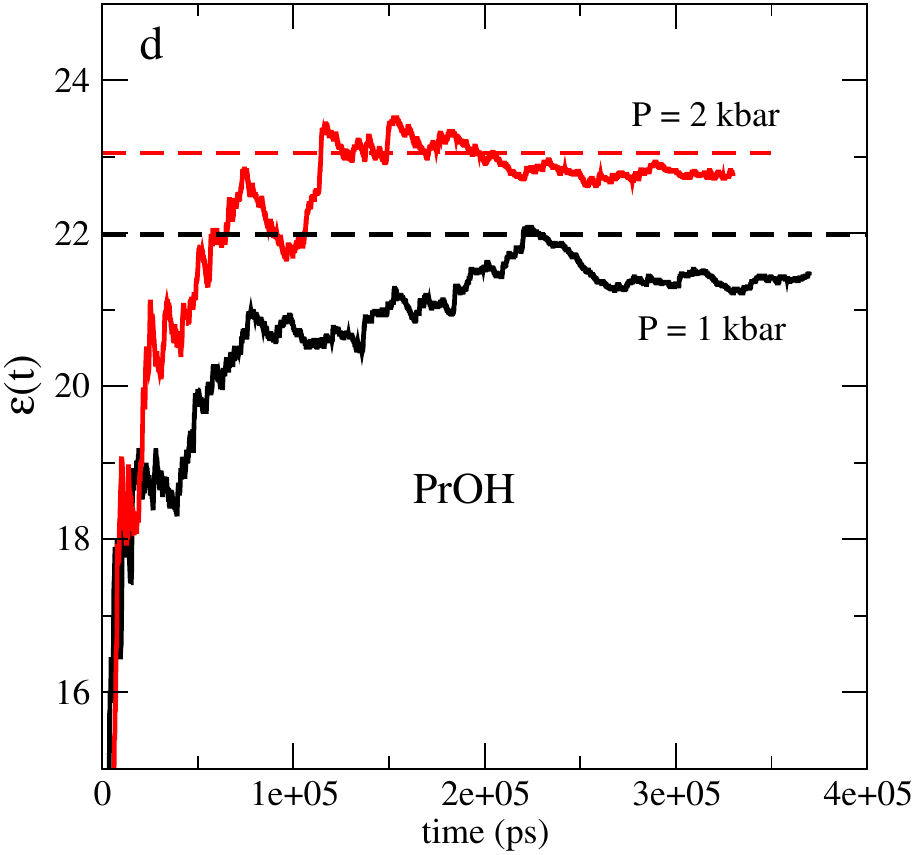}
\end{center}
\caption{(Colour online) Panels a, b, c and d: Illustration of the calculations of
dielectric constant of methanol, ethanol and 1-propanol
at different pressures (indicated in the figure) at 298.15~K.
The simulation results are for  UAM-EW united atom  model.
The experimental data (dashed lines in all panels) are from~\cite{moriyoshi}.
}
\label{fig3}
\end{figure}

Technically, the value for $\varepsilon$ is deduced from a
plateau region of the curves for $\varepsilon (t)$. A few examples 
illustrating the accuracy of our calculations for the three
alcohols under study are shown in figure~\ref{fig3}.
The experimental data are taken from~\cite{moriyoshi}.
Earlier results of measurements of $\varepsilon$ for three
alcohols are given in~\cite{kay}. They are in 
good agreement with \cite{moriyoshi} for MeOH and EtOH.
For PrOH, the qualitative behavior is similar, just the data
from \cite{kay} are very slightly lower than those from \cite{moriyoshi}.
Therefore, we restricted our comparisons to the more recent data set.
As expected, long runs (trajectories) are necessary to capture 
the value for the dielectric constant with reasonable accuracy.
In the case of MeOH and EtOH (panels a and b of figure~\ref{fig3}), 
the curves for $\varepsilon(t)$ exhibit plateaus after 200~ns approximately.
It can be seen that the UAM-EW model slightly underestimates the values for the dielectric
constant of MeOH in the entire interval of pressures under study (panel a of figure~\ref{fig3}).
On the other hand, for EtOH the model either underestimates $\varepsilon$ at
low pressures, close to ambient pressure, or overestimates it at high pressure values
(panel b, figure~\ref{fig3}). 

The accuracy of calculations for PrOH are illustrated in figures~\ref{fig3}c and ~\ref{fig3}d.
In some cases, the $\varepsilon(t)$ curves
exhibit more pronounced oscillations (note that the $y$-axis range is narrower here 
than it is in panels a and b) comparing to MeOH and EtOH.
Perhaps both the re-organization of the  non-polar part and the 
assumed rigidity of the molecules
contribute to the rate of decay of polarization fluctuations. 
Deeper understanding of these issues require extensive studies
of dielectric relaxation phenomena --- a wide and active area of research in spite
of a quite long history 
for various alcohols~\cite{saiz1,yao,noskov,ludwig1,bagchi,marx}.

\begin{figure}[h]
\begin{center}
\includegraphics[width=6.0cm,clip]{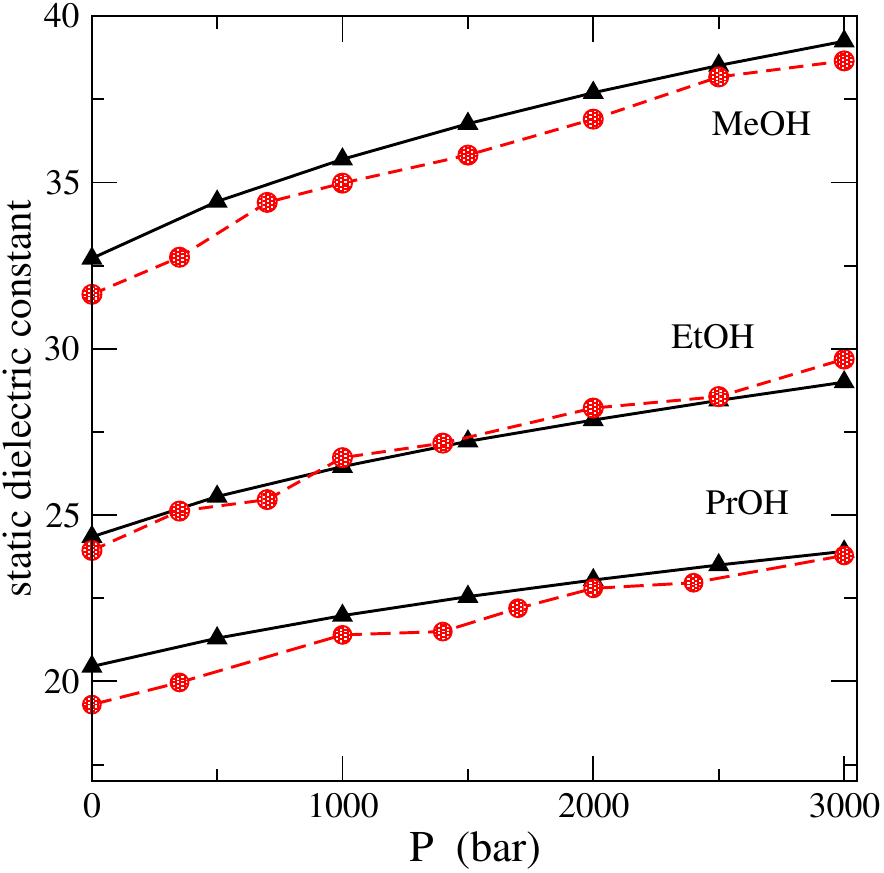}
\end{center}
\caption{(Colour online) Static
dielectric constant of alcohols
on pressure at 298.15~K.
The simulation results are for the UAM-EW united atom  model.
The experimental data (solid lines with triangles) are from \cite{moriyoshi}.
}
\label{fig4}
\end{figure}

A summarizing insight into the dependence of the dielectric constant on pressure
for the three alcohols under study is provided by figure~\ref{fig4}.  In general terms, the UAM-EW
model satisfactorily describes the trends of behavior of the static dielectric constant
on pressure starting from 1 bar up to 3~kbar. Discrepancy between predictions coming
from simulations and experimental data is of the order of a few percent.
The dielectric constant increases with increasing pressure for all three alcohols.

However, the increment is most pronounced for MeOH, in comparison with ETOH and PrOH. For
this latter system, the growth of $\varepsilon$ is the weakest. In fact, trends for density,
cf. figure~\ref{fig1}, and the dielectric constant with pressure are qualitatively similar.
Apparently, the correlations between dipolar moments of molecules become stronger
when the average distance between them decreases.
We will return to this point below, after discussing the changes of the microscopic structure 
with increasing pressure. 

\subsection{Self-diffusion coefficients}

Evaluation of the quality of the model frequently involves the results for the
self-diffusion coefficient, $D$. Note, that $D$ has not been considered  as a target
property within the multistep parametrization of the UAM-EW model~\cite{melgarejo}.
One of the common routes to obtain $D$,  is from the
mean square displacement of particles. On the other hand, it may be also calculated from the
velocity
auto-correlation function.
We calculate $D$  by the former procedure, via the Einstein relation,
\begin{equation}
D =\frac{1}{6} \lim_{t \rightarrow \infty} \frac{\rd}{\rd t} \vert {\bf r}(\tau+t)
-{\bf r}(\tau)\vert ^2,
\end{equation}
where  $\tau$ denotes the time origin. Default settings of GROMACS were used 
for the separation of the time origins.

\begin{figure}[h]
\begin{center}
\includegraphics[width=6.0cm,clip]{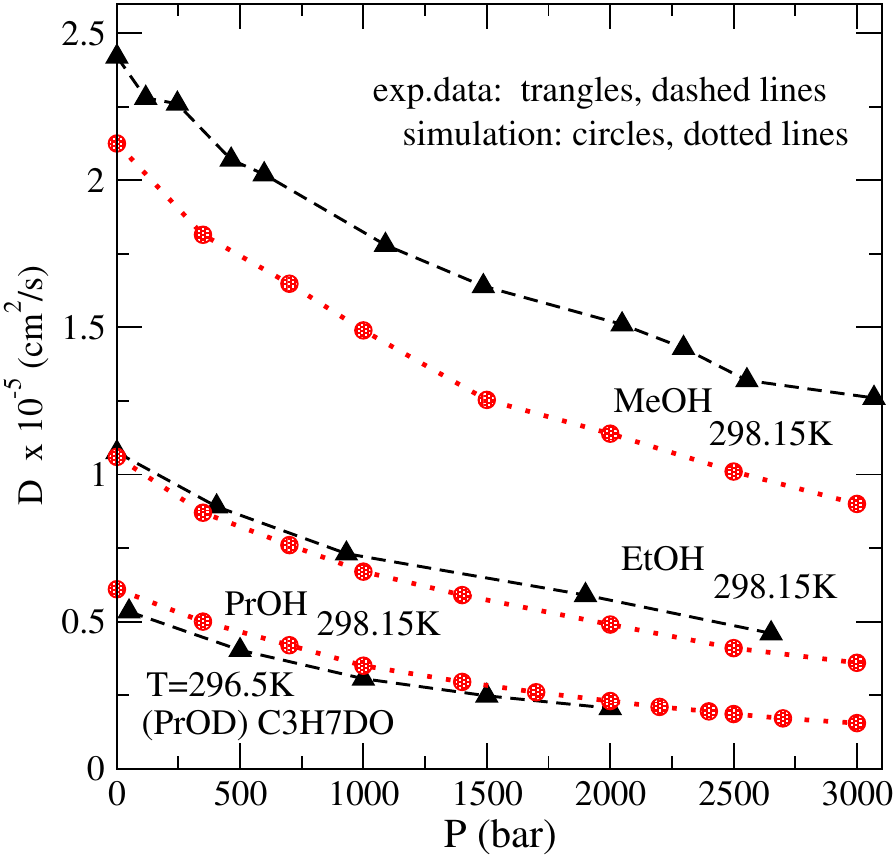}
\end{center}
\caption{(Colour online) Self-diffusion coefficient of alcohols
on pressure at 298.15~K.
The simulation results (circles and dotted lines) are for the UAM-EW united atom
model at 298.15~K.
The experimental data (dashed lines with triangles) are from Hurle et al.~\cite{hurle}
(methanol and ethanol at 298.15~K).
The experimental data for PrOH are for a C3H7DO sample at 296.5~K
from Shaker-Goafar et al.~\cite{karger}.
}
\label{fig5}
\end{figure}

The dependence of $D$ on pressure following from simulations, for the three alcohols in question,
is shown in figure~\ref{fig5}. Besides, this figure contains available experimental data
for the sake of comparison. It can be seen that $D$ decreases with increasing
pressure for each alcohols in agreement with experimental data. 
This kind of behavior is expected due to augmenting density with 
pressure. 

For MeOH, the decrement of the self-diffusion coefficient with pressure is larger than
for the two other alcohols, EtOH and PrOH, as seen from experimental data and simulations.
The values for $D$ for MeOH are underestimated over the entire pressure range
compared to experimental results. On the other hand, the agreement between simulations
and experiments is much better for EtOH and PrOH.

Undoubtedly, it is worth complementing this type of results 
for alcohols under pressure by the calculations of viscosity and other 
dynamic properties as it was discussed in \cite{zeidler}.
We hope to extend the present study in this aspect in a future work.

\subsection{Microscopic structure}

This subsection contains the site-site pair distribution functions (PDFs) for different 
alcohols and some coordination numbers necessary to interpret microscopic structure.

\begin{figure}[h]
\begin{center}
\includegraphics[width=5.5cm,clip]{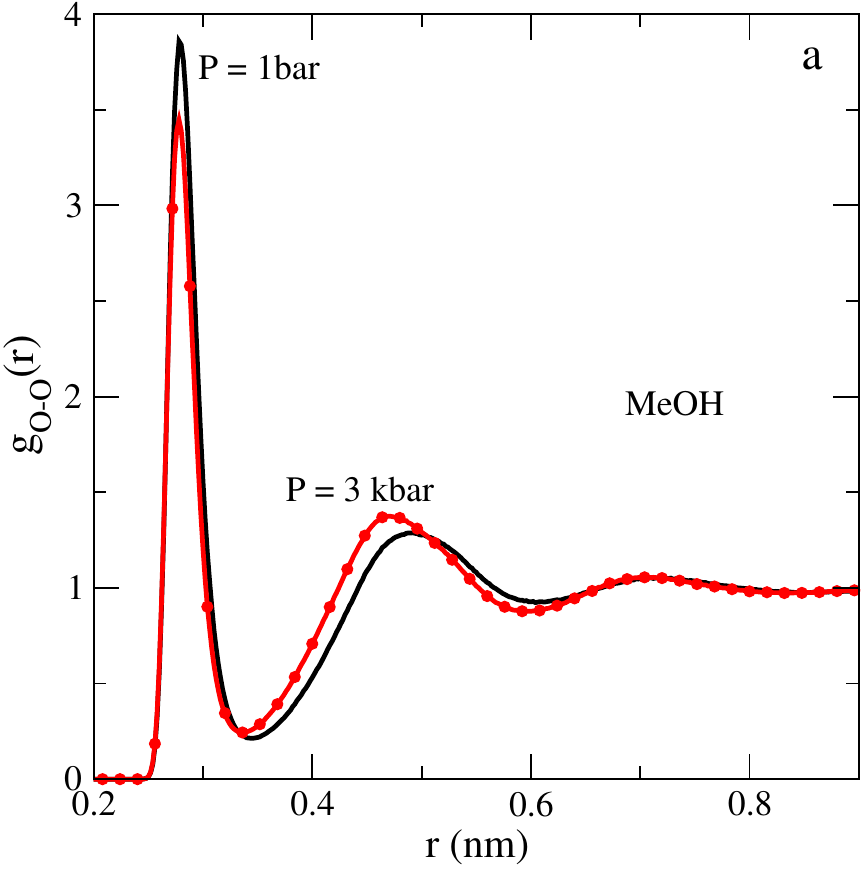}
\includegraphics[width=5.5cm,clip]{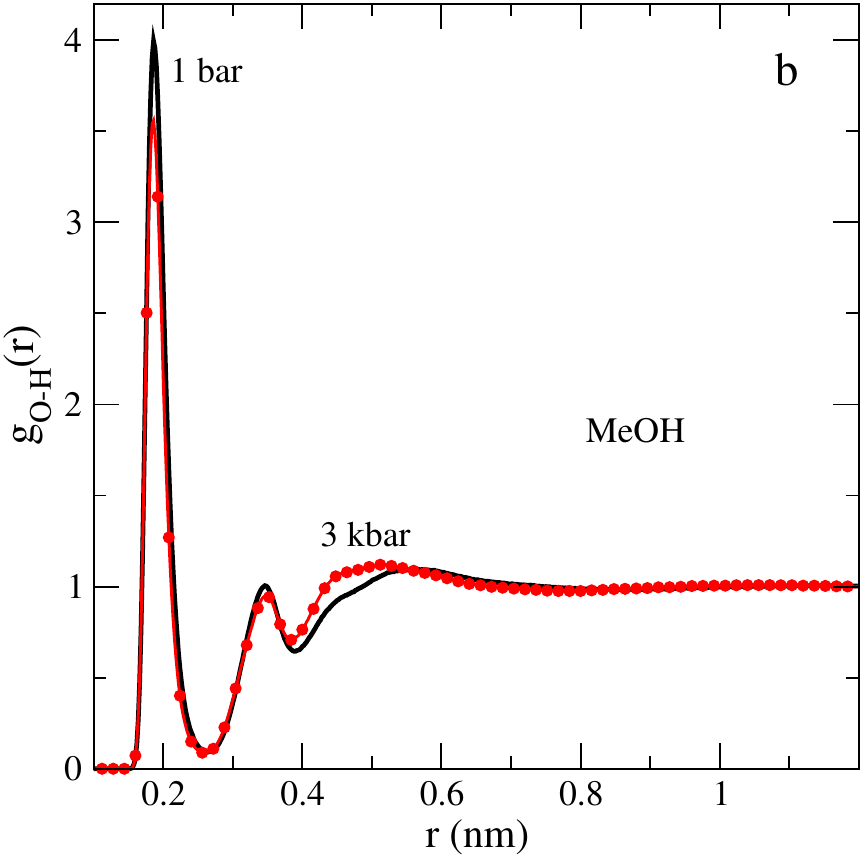}
\includegraphics[width=5.5cm,clip]{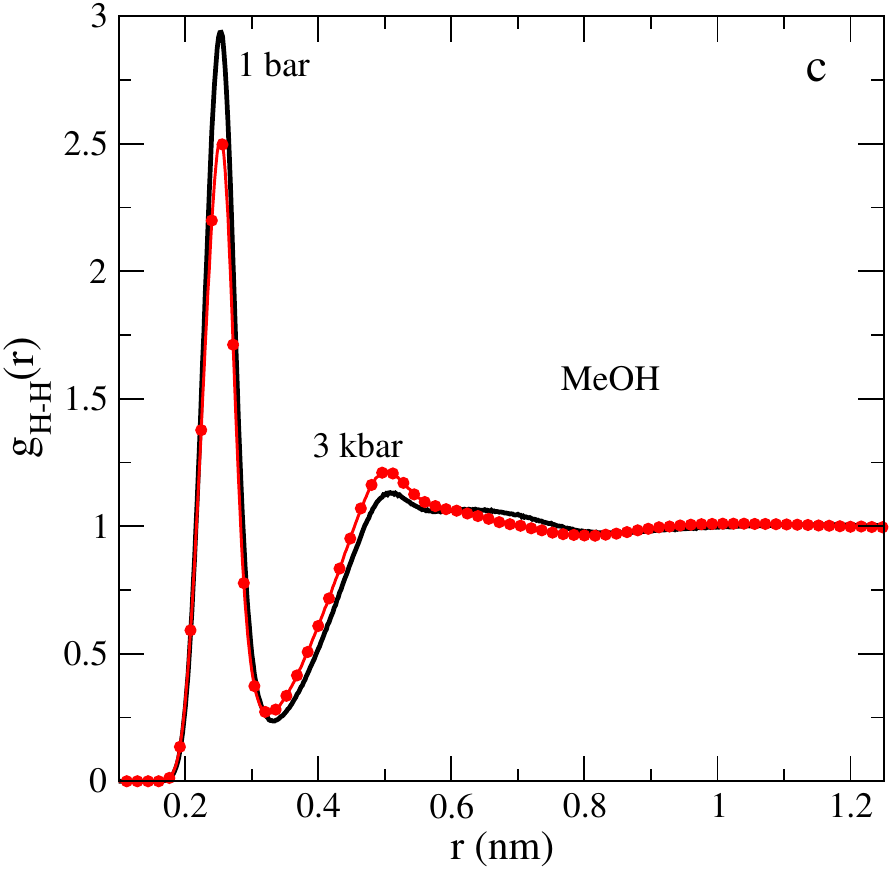}
\end{center}
\caption{(Colour online) Computer simulation results for the site-site
pair distribution functions for liquid MeOH at 1 bar (black lines) and
at 3 kbar (red lines) at 298.15~K.
}
\label{fig6}
\end{figure}

\begin{figure}[h]
\begin{center}
\includegraphics[width=5.3cm,clip]{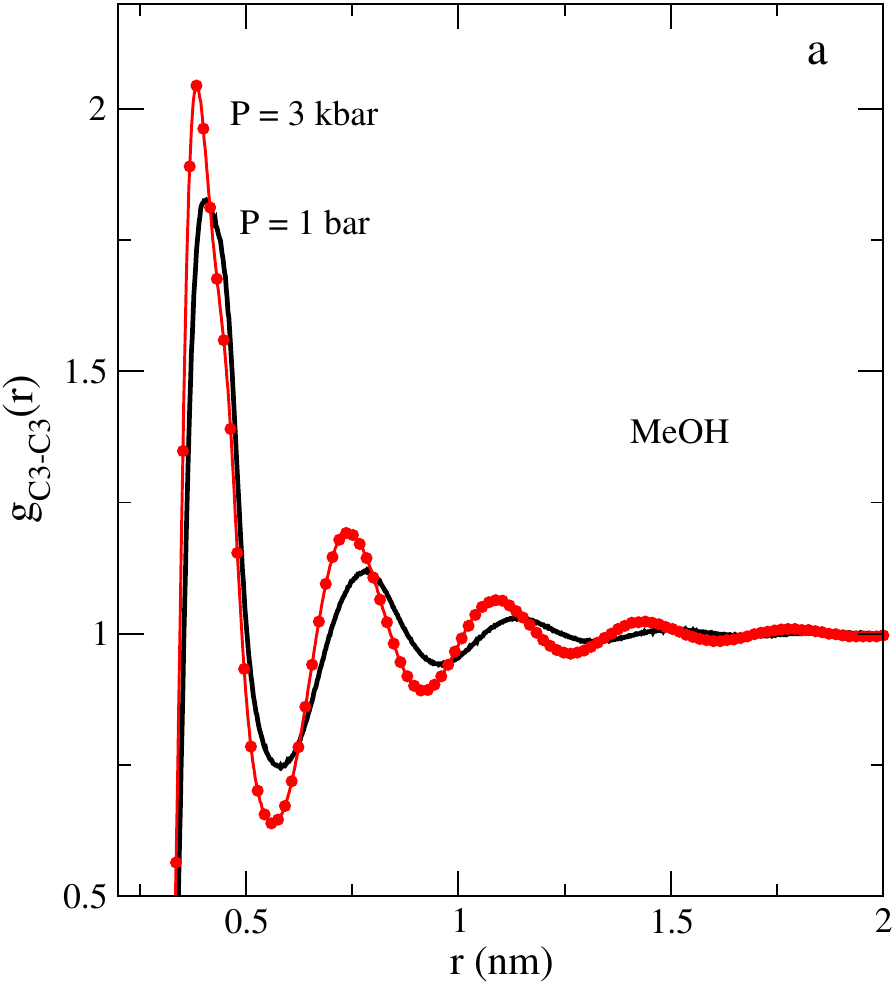}
\includegraphics[width=6cm,clip]{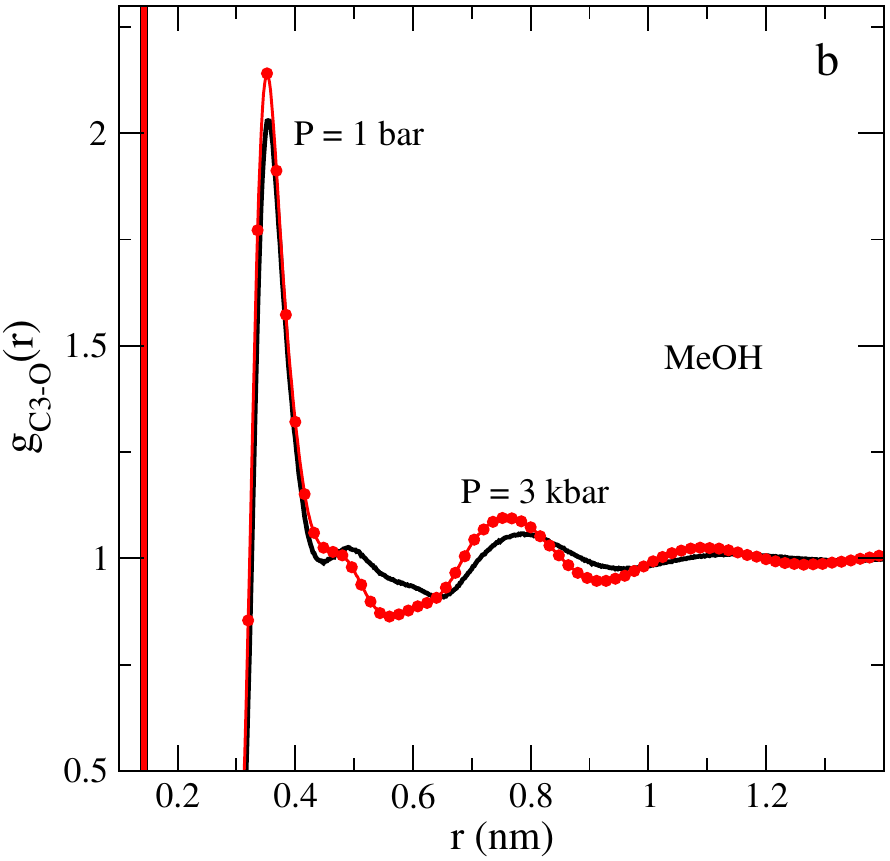}
\end{center}
\caption{(Colour online) Computer simulation results for the site-site
pair distribution functions for liquid MeOH at 1 bar (black lines) and
at 3 kbar (red lines) at 298.15K.
}
\label{fig7}
\end{figure}

Concerning the first maxima of the functions shown in figure~\ref{fig6} for MeOH,
we would like to mention the following features.
The height of the first maximum of $g_{\text{O-O}}(r)$  and $g_{\text{O-H}}(r)$, panels a and b, respectively,
slightly decreases in value, if the pressure increases from 1 bar to 3 kbar.
By contrast, the first maximum of $g_{\text{C3-C3}}(r)$  and $g_{\text{C3-O}}(r)$ (figure~\ref{fig7}),
increases upon increasing pressure. The first and second coordination shells are well
pronounced in the  $g_{\text{O-O}}(r)$ PDF at 1 bar and at 3 kbar, figure~\ref{fig6}a. 
The third shell is weakly pronounced in both cases. 

The maxima characterizing the first
shell centered on an oxygen atom in terms of $g_{\text{O-O}}(r)$  and $g_{\text{O-H}}(r)$, witness 
that the distances between these atoms do not change. Thus, the first shell remains
practically intact upon increasing pressure. However, the characteristic distances
describing more distant atoms change. Namely, the second maximum of $g_{\text{O-O}}(r)$
shifts to a shorter distance at 3 kbar compared to 1~bar (figure~\ref{fig6}a). 
Changes of the height of the first and the second minima of O-O distribution
are small. This behavior confirms a weak distortion of the neighborhood of
oxygen atoms and of the arrangement of  polar parts of methanol molecules.
The $g_{\text{H-H}}(r)$ PDF keeps unchanged its principal features with increasing pressure
as well, figure~\ref{fig6}c.
Interpretation of the structure in terms of PDFs for MeOH was elaborated
in several previous publications. We would like to attract attention of the reader to 
the results from \cite{weitkamp} and our recent findings~\cite{pusztai1}.

Additional insights into the changes of the O-H coordination with pressure appear from table~\ref{TableI}.
The first and following coordination numbers are commonly evaluated  by integration
of the pair distribution functions up to the first and following minima,
\begin{equation} 
  n_{ij} = 4\piup \rho_j \int_0 ^{r^{\text{min}}_{ij}}  g_{ij}(R)  R^2 \rd R,
\end{equation}
where $\rho_j$ is the density of species $j$ and  $g_{ij}(r)$ is the 
appropriate PDF. The first coordination number practically does not change
with pressure ($n_{\text{OH}}^{(i)} = n^{(i)}$). 
However, the second coordination number increases, presumably
due to increasing density of the system at a higher pressure, 3 kbar.
In addition  to the data in table~\ref{TableI}, we have calculated the
average number of hydrogen bonds, $n_{\text{HB}}$, per methanol molecule.
It is obtained by using the GROMACS utility with default settings, i.e.,
the geometric criteria are used. At 1 bar, $n_{\text{HB}}$ = 1.923, whereas at
3 kbar, $n_{\text{HB}}$ = 1.9355. In conclusion, pressure affects the
number of hydrogen bonds between the neighboring molecules quite weakly.
This observation is in accordance with our very recent extended discussion
in \cite{pusztai1} that underlines changes of the cooperative structure
of hydrogen bonding network, rather than  the influence of pressure on $n_{\text{HB}}$.

\begin{table}[h]
  \centering
   \caption{
  Location of the first two minima of O-H distribution,
	and the corresponding coordination numbers, $n_{\text{OH}}^{(1)}$
	and $n_{\text{OH}}^{(2)}$, for MeOH at two pressures.
   \vspace{0.1cm}
   }
   \begin{tabular}{ccccc
   }
  \hline
	   $P$  & $r_{\text{min}}^{(1)}$ (nm) &  $n^{(1)}$  & $r_{\text{min}}^{(2)}$ (nm) & $n^{(2)}$
      \\[0.5ex]
     \hline
1 bar & 0.264 & 1.985 & 0.389 & 3.596  \\
3 kbar & 0.259 & 1.991 & 0.384 & 3.797  \\
\hline
\end{tabular}
\label{TableI}
\end{table}

Much more pronounced changes of the structure upon increasing
pressure are observed in the function $g_{\text{C3-C3}}(r)$, figure~\ref{fig7}a. Its oscillations 
essentially grow in magnitude. At least four coordination shells
are seen that witness the trends to form a more ordered structure of non-polar
``tails'' of methanol molecules. This kind of behavior was 
observed previously, cf. panels a and b of figure~\ref{fig4} in the \cite{pusztai1}.
In addition, one can find a very satisfactory agreement of the characteristic
distances for this PDF and the following $g_{\text{C3-O}}(r)$ (figure~\ref{fig7}b), both
with the curves shown in figure~\ref{fig4} from \cite{pusztai1} and figure~\ref{fig5} in
\cite{weitkamp}.

\begin{figure}[h!]
\begin{center}
\includegraphics[width=4.7cm,clip]{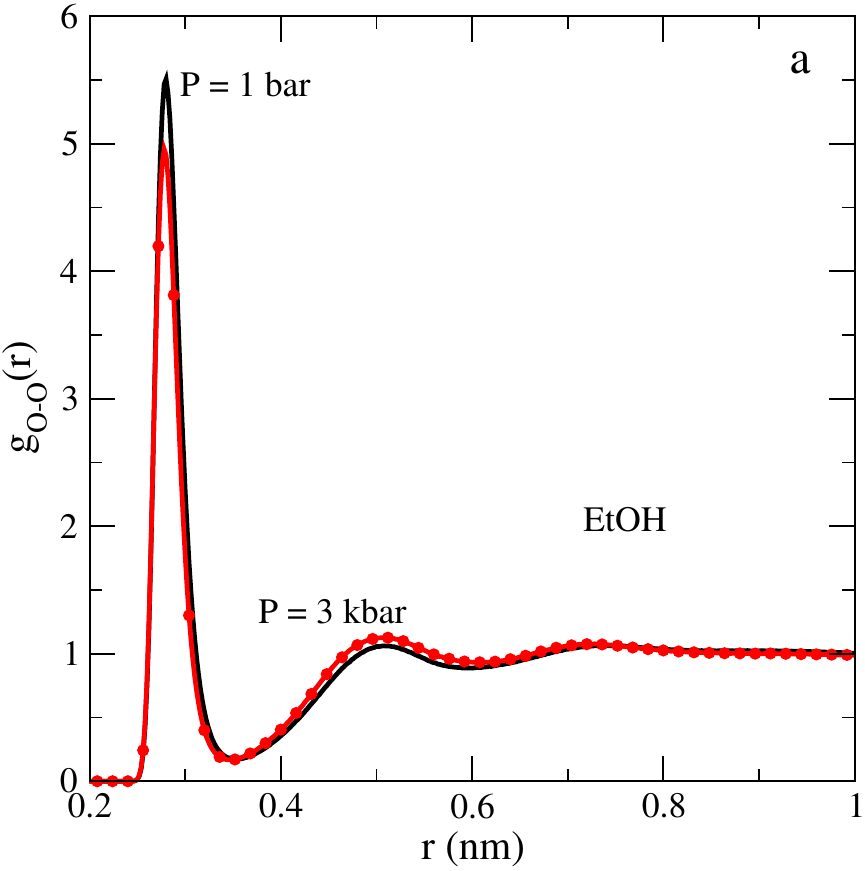}
\includegraphics[width=4.7cm,clip]{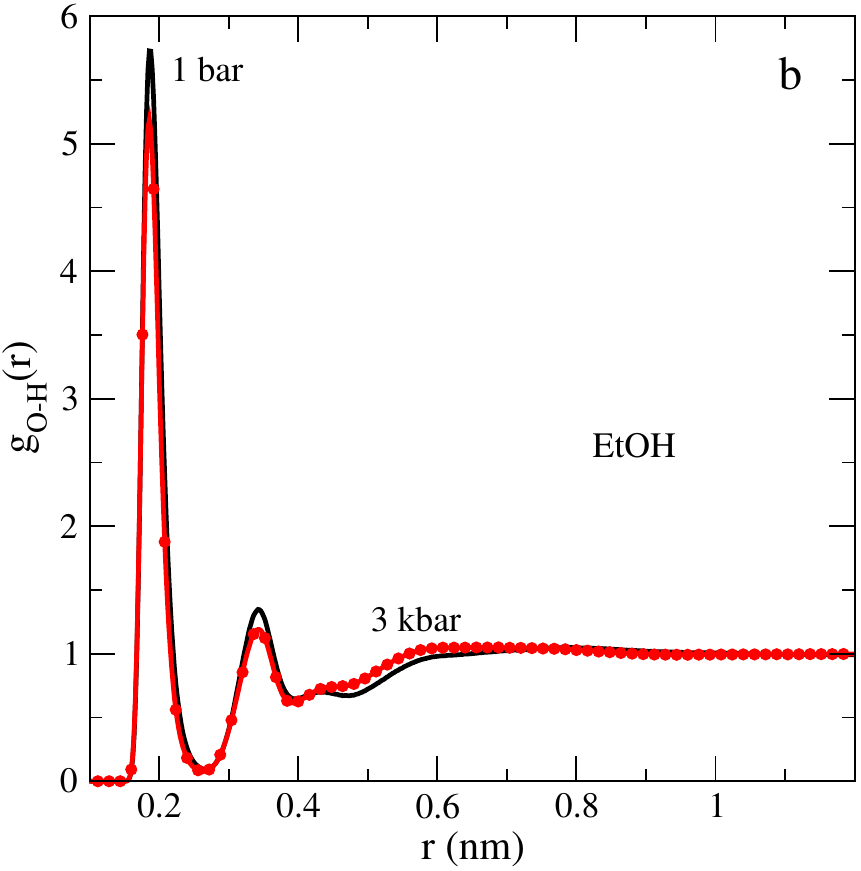} \\
\includegraphics[width=4.7cm,clip]{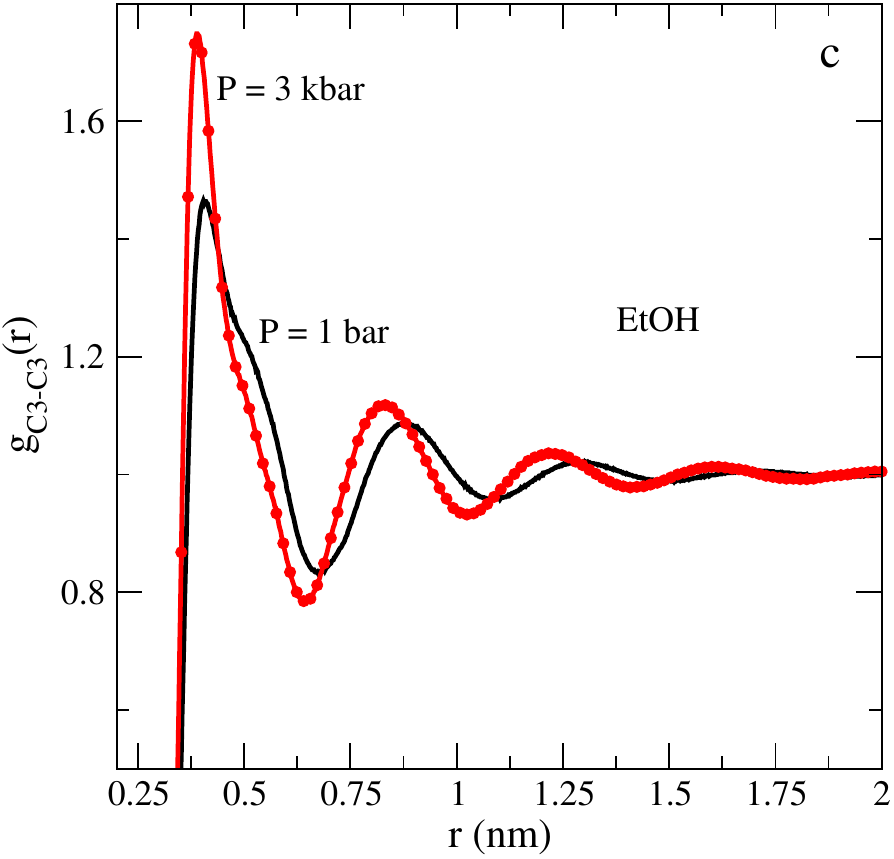}
\includegraphics[width=4.7cm,clip]{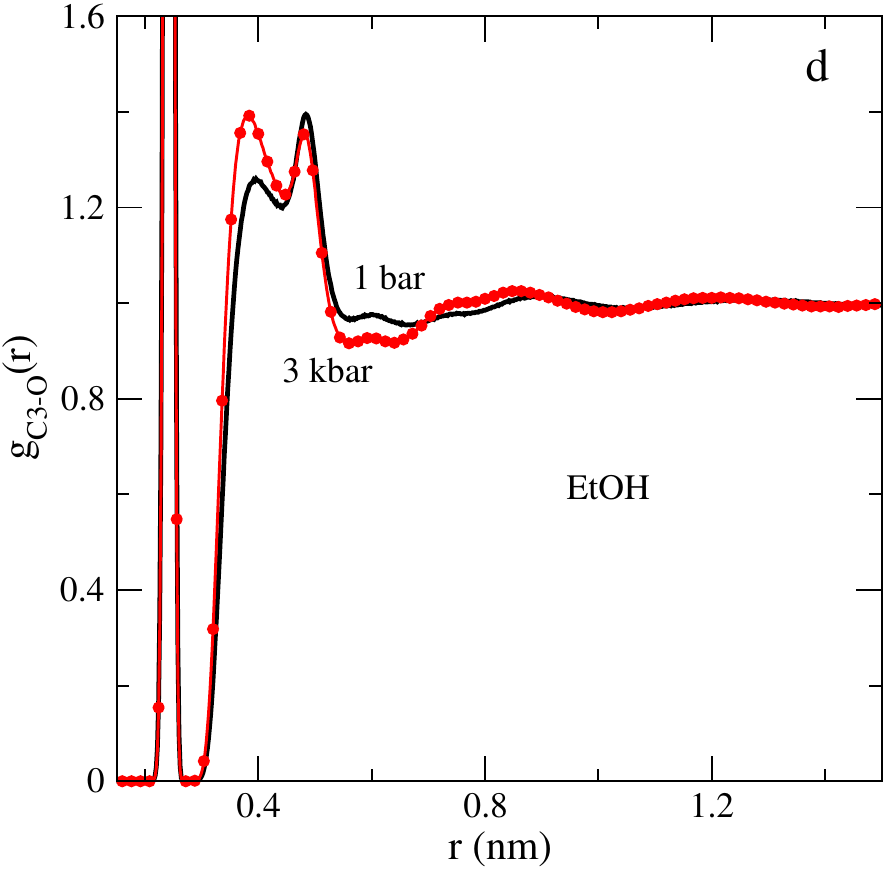}
\end{center}
\caption{(Colour online)
Computer simulation results for the site-site
pair distribution functions for liquid EtOH at 1 bar (black lines) and
at 3 kbar (red lines) at 298.15~K.
}
\label{fig8}
\end{figure}

Next, we proceed to the EtOH system and describe changes of its  structure
with increasing pressure in figure~\ref{fig8}. The first two panels,  a and b,
concern the polar part of EtOH molecule, figures~\ref{fig8}a and \ref{fig8}b.
Changes of the first coordination shell in this system in terms of the functions
$g_{\text{O-O}}(r)$  and $g_{\text{O-H}}(r)$ are marginal, in close similarity 
to MeOH discussed above. However, the shape of the 
$g_{\text{O-H}}(r)$ PDF differs from what was observed for MeOH. Namely,
there is a well pronounced second maximum on O-H distribution
as an evidence of a ``tighter'' arrangement of the second coordination
shell, in contrast to fluid MeOH.
A more complex shape of $g_{\text{C3-C3}}(r)$  and $g_{\text{C3-O}}(r)$ PDFs (figures~\ref{fig8}c and \ref{fig8}d), in
comparison with MeOH, is due to the two-site non-polar part of the EtOH molecule
comprising CH$_2$ and CH$_3$ sites. Moreover, the magnitude of changes of the
shape of $g_{\text{C3-C3}}(r)$  and $g_{\text{C3-O}}(r)$ PDFs upon increasing pressure,
are more pronounced in comparison with MeOH.  Unfortunately, we are not able
to perform detailed comparisons of trends for the microscopic structure
upon increasing pressure. The experimental data at high pressure are
not available in the literature so far. Apparently, principal features
of the structure predicted by the model in question qualitatively agree
with findings discussed in \cite{vrhovsek}. We expect that present
findings will stimulate experimental research at high pressures for
neat EtOH and its aqueous solutions, similarly to 
the MeOH system \cite{pusztai2}.

\begin{figure}[h]
\begin{center}
\includegraphics[width=5cm,clip]{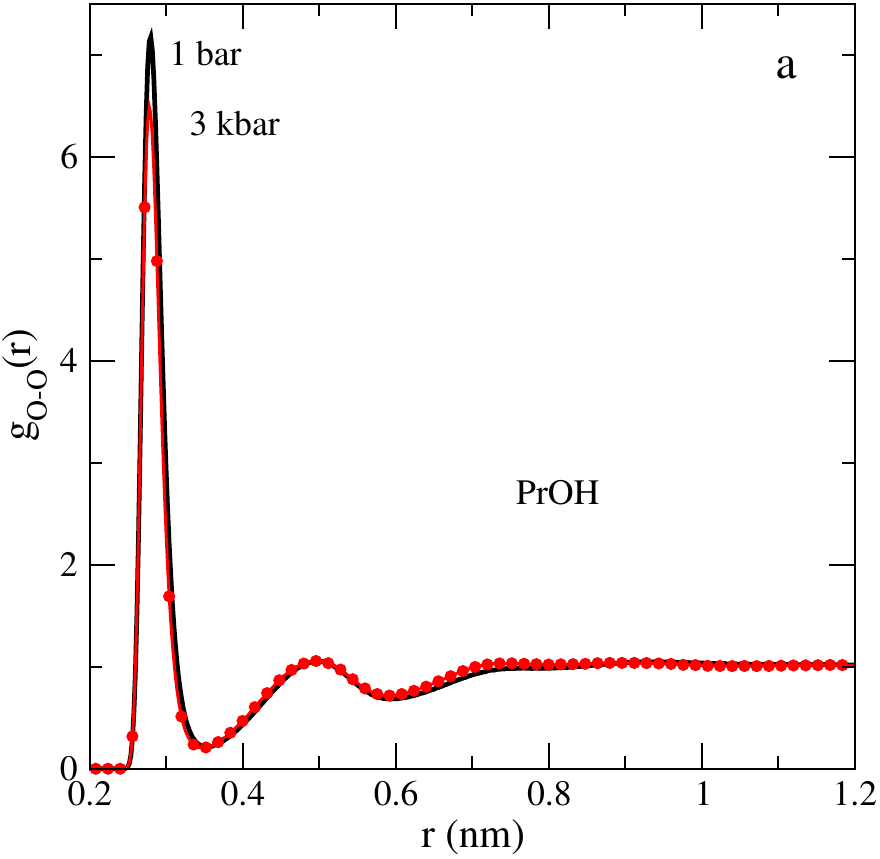}
\includegraphics[width=5cm,clip]{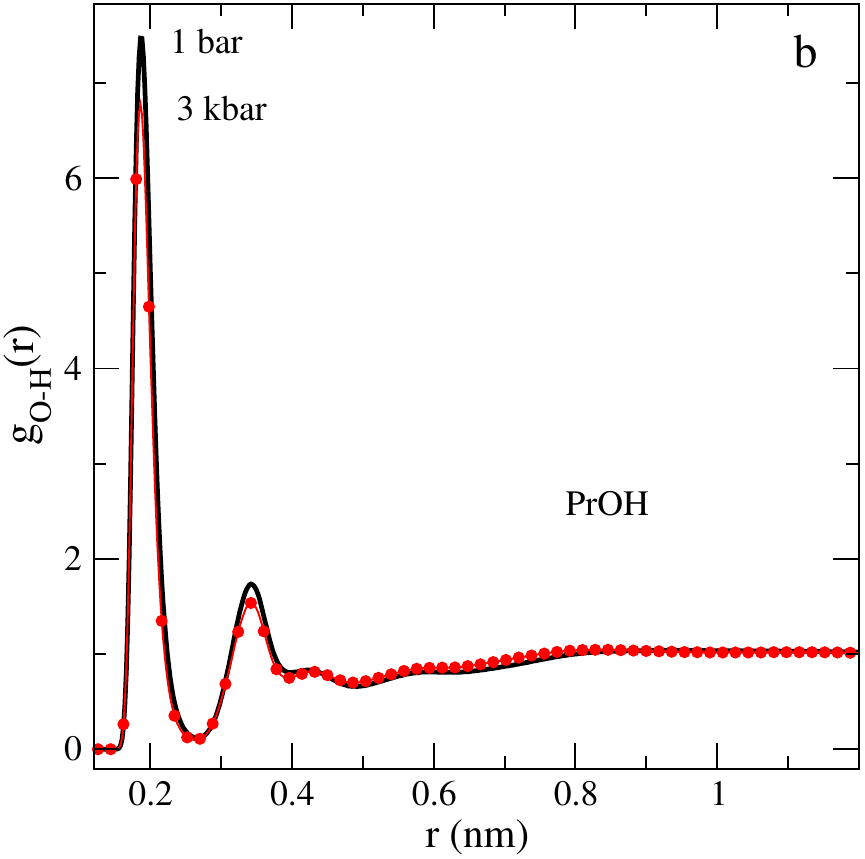}
\end{center}
\caption{(Colour online)
Computer simulation results for the site-site
pair distribution functions for liquid PrOH at 1 bar (black lines) and
at 3 kbar (red lines) at 298.15~K.
}
\label{fig9}
\end{figure}

\begin{figure}[h]
\begin{center}
\includegraphics[width=5.5cm,clip]{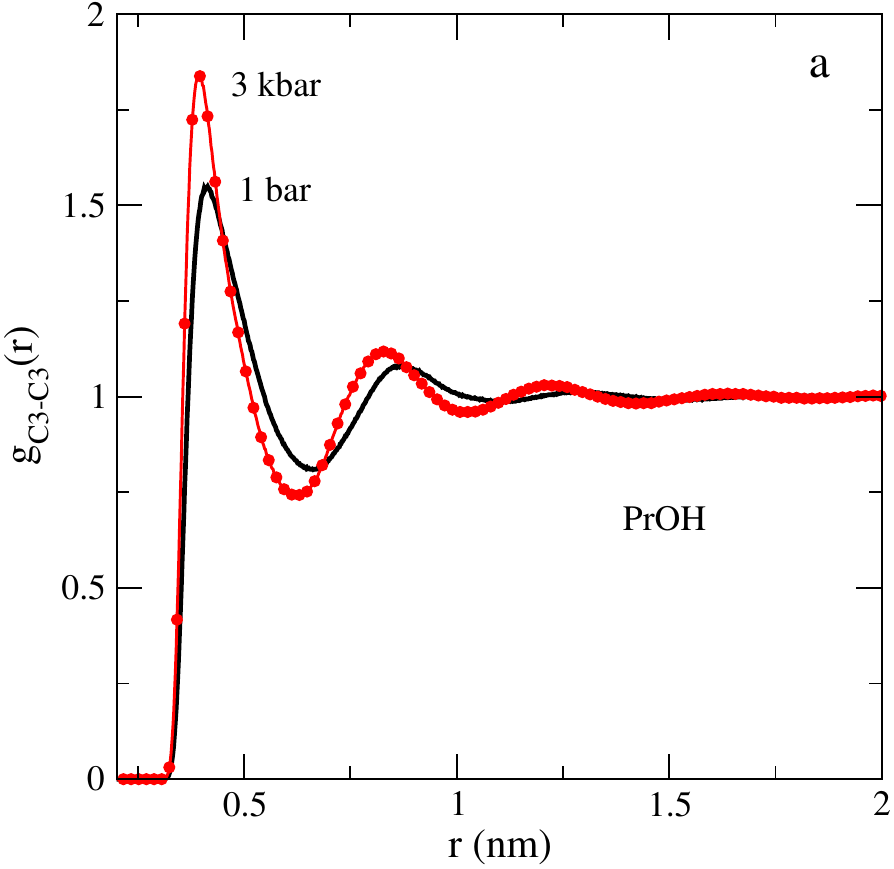}
\includegraphics[width=5.5cm,clip]{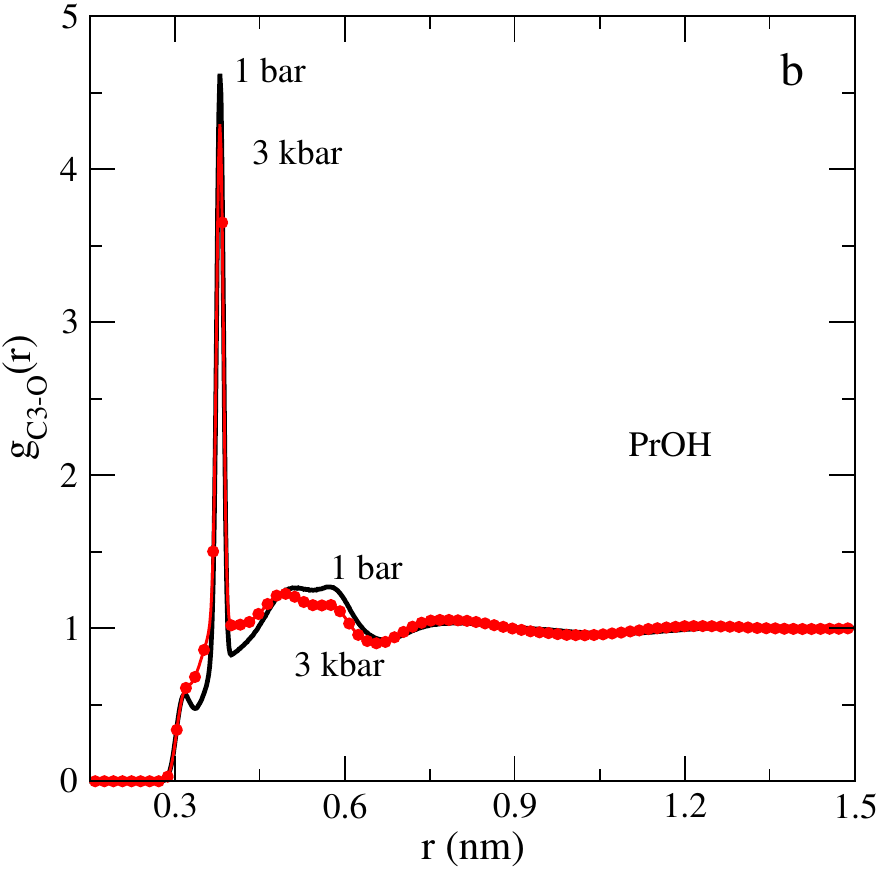}
\end{center}
\caption{(Colour online)
Computer simulation results for the site-site
pair distribution functions for liquid PrOH at 1 bar (black lines) and
at 3 kbar (red lines) at 298.15~K.
}
\label{fig10}
\end{figure}

The effect of pressure on the microscopic structure of PrOH is described
in figures~\ref{fig9} and \ref{fig10}.
Again, we resort to the evolution of the distribution functions
and some coordination numbers. The principal findings can be summarized as 
follows.  The functions describing the distribution of atoms of the polar part of
the molecule, O and H (panels a and b, figure~\ref{fig9}), respond very weakly to augmenting pressure.

\begin{table}[h]
  \centering
   \caption{
  Location of the first minima of O-H distribution,
and the corresponding coordination numbers for PrOH
 at two pressures. The distances marked as $2a$ and $2b$ correspond to a minimum
 on the shoulder of the second shell and to the true minimum between the second
	and third shell in figure~\ref{fig9}b.
   \vspace{0.1cm}
   }
   \begin{tabular}{ccccccc
   }
  \hline
	   $P$  & $r_{\text{min}}^{(1)}$ (nm) &  $n^{(1)}$  & $r_{\text{min}}^{(2a)}$ (nm) & $n^{(2a)}$
	   & $r_{\text{min}}^{(2b)}$ (nm) & $n^{(2b)}$
      \\[0.5ex]
     \hline
           1 bar & 0.267 & 1.987 & 0.397 & 3.472  & 0.492 & 4.853  \\
           3 kbar & 0.263 & 1.991 & 0.395 & 3.535 & 0.486 & 5.115  \\
\hline
\end{tabular}
\label{TableII}
\end{table}

The average number of hydrogen bonds per propanol molecule, $n_{\text{HB}}$,
at 1 bar is 1.937, whereas at 3~kbar it is 1.9512. These numbers
at two pressures are close to the first coordination numbers in table II
and confirm that the majority of molecules forming the first shell are
hydrogen bonded.
Moreover, the bonding between the neighboring molecules
is marginally affected by the application of pressure on the system.

The structure of the first coordination shell of oxygens remains practically
the same at 1 bar and at 3 kbar. Only farther shells slightly change, as
manifested in the $g_{\text{O-H}}(r)$ distribution (figure~\ref{fig9}b), due to the changes of arrangement
of the non-polar parts of molecules (figure~\ref{fig10}). These non-polar tails of PrOH particles,
composed of two CH2 and one CH3 sites, noticeably respond to compression of the system figure~\ref{fig10}a. 
In contrast to figure~\ref{fig8}d for EtOH, where the intra-
and inter-molecular contributions are separated, 
for PrOH,  we cannot separate these terms, i.e., they overlap
due to a larger non-polar tail, figure~\ref{fig10}b. This issue does not permit to straightforwardly
interpret the changes of $g_{\text{C3-O}}(r)$ on pressure.

Our final remarks in this section concern the behavior of the 
distance dependent finite system Kirkwood factor, $G_\text{K}$.
The theoretical background of the significance of $G_\text{K}$ for dipolar hard 
spheres was explained in \cite{holovko} very much in detail.
On the other hand, we would like to refer to \cite{ortiz1,ortiz2,boda}
that contain an appealing description of implications 
for more complex molecules and for fluids with hydrogen bonding.
In the present simulations, $G_\text{K}$, is calculated similarly to~\cite{salgado},

\begin{equation}
G_\text{K} = 1 + \frac{2}{N} \sum_{i<j} \mathbf u_i  \mathbf u_j ,
\end{equation}
where $N$ is the number of molecules in the box,
$\mathbf{u}_{i}$ is the unit vector in the direction of the dipole moment 
of molecule $i$.

\begin{figure}[h]
\begin{center}
\includegraphics[width=5.5cm,clip]{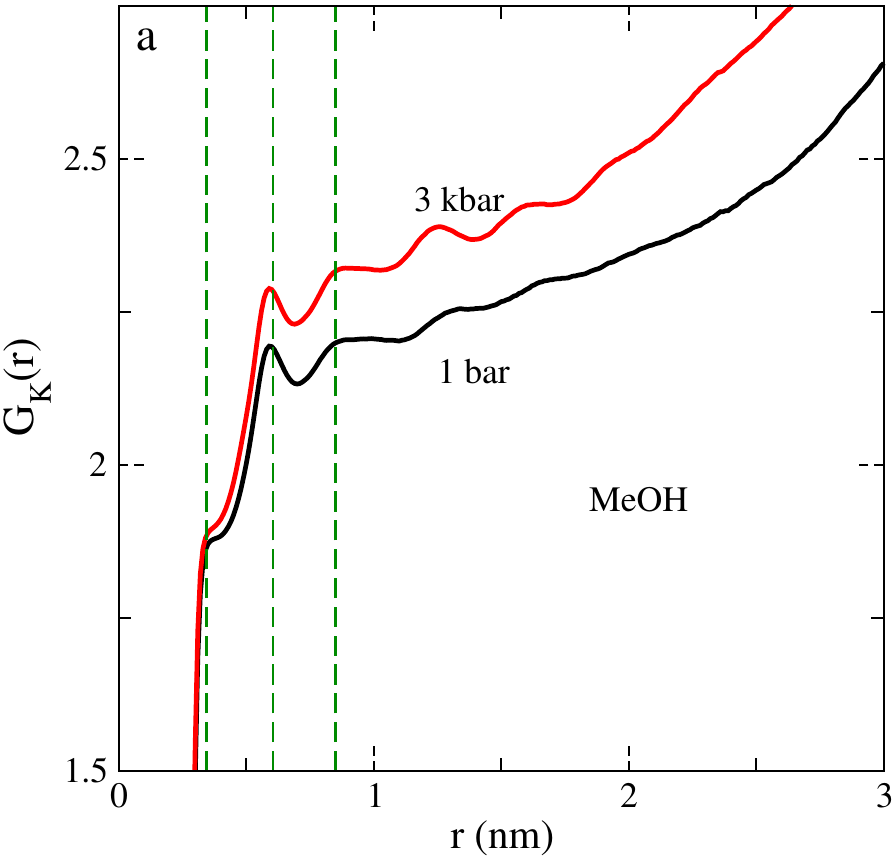}
\includegraphics[width=5.5cm,clip]{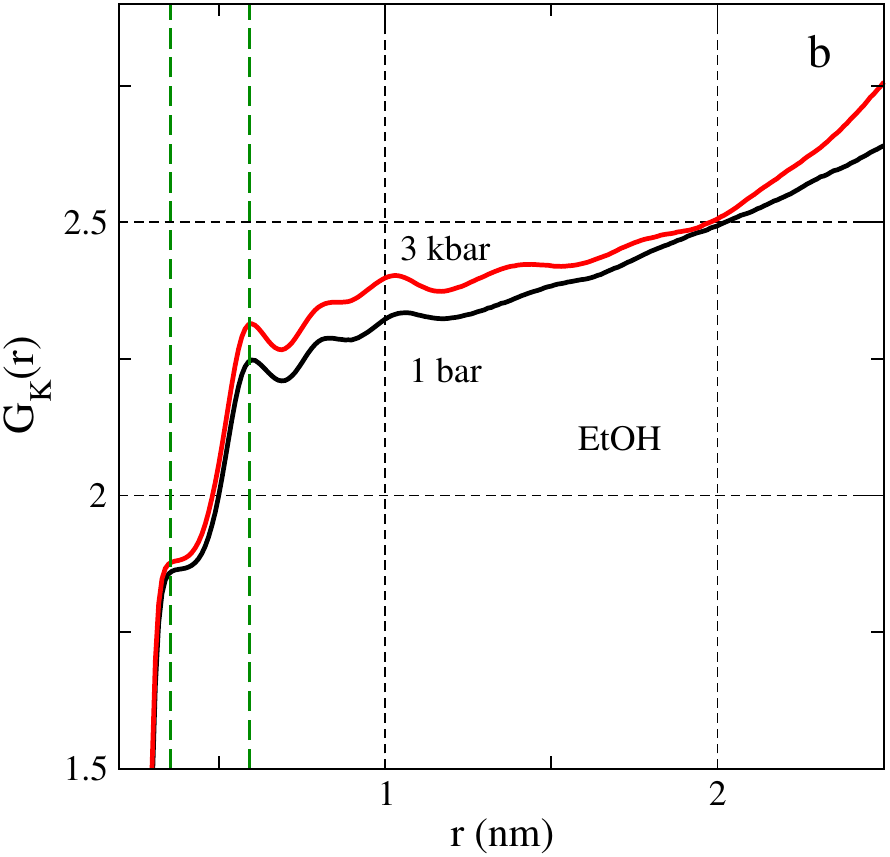}
\includegraphics[width=5.5cm,clip]{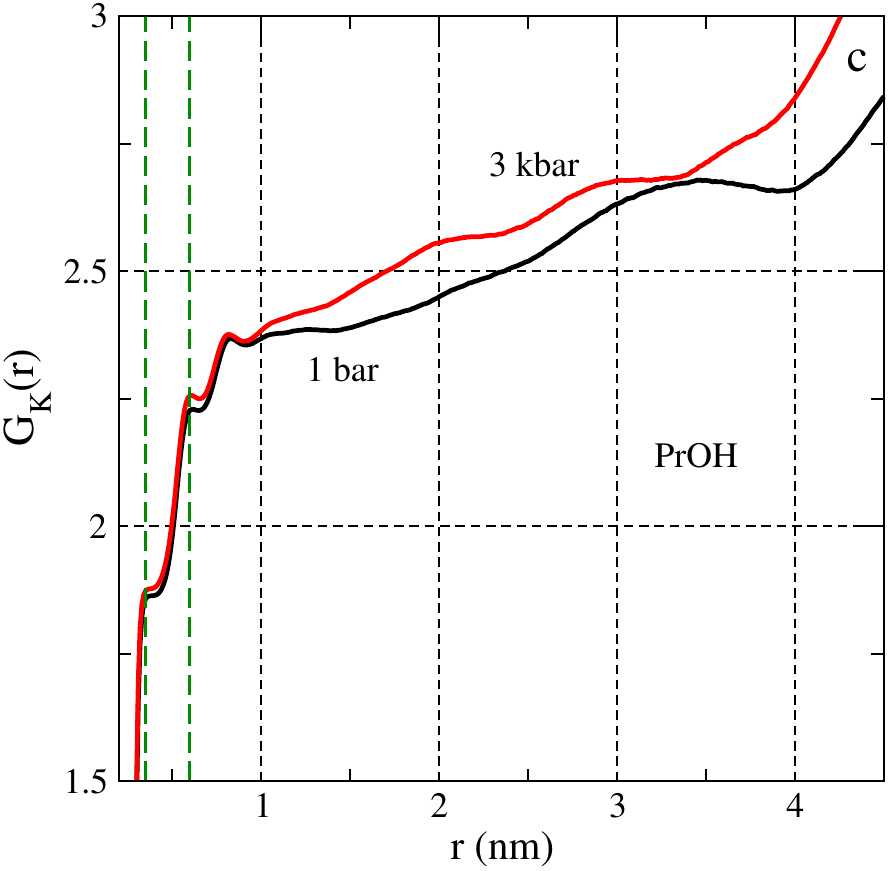}
\end{center}
\caption{(Colour online) Illustration of the calculations of
the distance dependent finite --- system Kirkwood factor, $G_\text{K}(r)$,
$r$ is the distance between the centres of two dipoles. The results for
MeOH, EtOH and PrOH are given in  panels  a, b and c, respectively.
The pressures are indicated in the figure, $T=298.15$~K.
}
\label{fig11}
\end{figure}

Some issues concerning the models for MeOH were considered in
\cite{ortiz1,ortiz2,salgado} at $P = 1$ bar. Our results for MeOH, EtOH and PrOH, at
$P = 1$ bar and $P =3$ kbar ($T = 298.15$~K) are shown in three panels of 
figure~\ref{fig11}, respectively.  In each of the panels, the vertical dashed green 
lines mark the limits of the first few coordination shells evaluated from the
respective minima of the $g_{\text{O-O}}(r)$ pair distribution function
(from the curves in figures \ref{fig6}a, \ref{fig8}a, \ref{fig9}a and \ref{fig10}a).

The following conclusions may be formulated concerning $G_\text{K}(r)$. 
The first few  maxima 
on $G_\text{K}(r)$ correspond precisely to the coordination shells
determined by the microscopic structure given in terms of $g_{\text{O-O}}(r)$.
These indicate preferential parallel alignment of the dipole moments of
molecules for each system under study. Nevertheless, the oscillations 
of $G_\text{K}(r)$ are pronounced at larger distances where one cannot 
identify the maxima and minima of the O-O distribution. This behavior
can be attributed to a long-range dipole-dipole interaction and to
the cooperative features of the hydrogen bonding in the alcohols studied.
The length of non-polar part of a molecule, i.e., the number of sites, 
influences the shape of $G_\text{K}(r)$. For example, it behaves differently
at distances $0.5\,\text{nm} < r < 0.9\, \text{nm}$ for MeOH and PrOH, figures~\ref{fig11}a and \ref{fig11}c. 
One can attribute the different behavior of
$G_\text{K}(r)$ to the shape of $g_{\text{O-O}}(r)$, cf. figure~\ref{fig10}a. 
The effect of pressure on the values of $G_\text{K}(r)$ is most pronounced
for MeOH, in  comparison with the other two alcohols, EtOH and PrOH.
Nevertheless, one can see that the effect of pressure is negligible 
at small distances between molecules that correspond to the first
coordination shell. Apparently, higher values of $G_\text{K}(r)$ at 
high pressure, 3 kbar, can be attributed to the evolution of
cooperative features of hydrogen bonding, because at a ``pair level'',
the average number of hydrogen bonds per molecule is not
affected by increasing pressure from 1 bar to 3 kbar.
These issues require additional exploration by using simulations 
and by experimental research.

\section{Summary and conclusions}

To summarize, in the present work we presented  computer
simulation results concerning the effects of pressure on some properties of 
three simple monohydric alcohols, MeOH, EtOH and ProOH, by using the united atom type
UAM-EW model. The pressure values are in the interval
from 1 bar to 3 kbar. A single value of temperature, $T = 298.15$~K was considered. The dependencies of density, isothermal compressibility,
dielectric constant and self-diffusion coefficient on pressure
were evaluated and compared with experimental data.
The model under study yields  the results that reasonably well agree with experimental
findings. Simulation results are novel and 
demonstrate validity of the model. 

In addition, we illustrated the changes of structural properties of the model for alcohols 
under pressure, by using site-site pair distribution functions. Their behavior is discussed
in terms of coordination numbers. It is shown that the 
behavior of the finite system Kirkwood factor is related to the coordination shells
of the distribution of oxygen sites at ambient and at high pressure.

Undoubtedly, the results of the present study can serve as a guide to complement
exploration of the model in various aspects. Namely, one should attempt to
describe the adiabatic compressibility and involve experimental measurements 
of the speed of sound, for the sake of comparison. Heat capacity at constant
pressure should be explored as well. Concerning the dynamic properties, 
future research must involve simulations of viscosity. It is of great interest
to complement the description of trends
of dielectric constant on pressure from the present work, by the 
study of dielectric relaxation phenomena. All these issues permit
evaluation of computer simulation modelling  w.r.t. experimental measurements
of thermodynamic and dynamic properties and application of dielectric spectroscopy.
Along these lines, one would be able to critically evaluate the validity of the
model and attempt construction of a more accurate force field.

Much work is required to improve the present understanding of the structural 
properties. Evidently, there are no tools to establish accuracy of the
partial (site-site) pair distribution functions. The quality of the
entire set of PDFs is commonly evaluated by
construction of the total structure factor, $S(Q)$, and
comparisons with the results from X-ray and neutron
diffraction experimental data at different conditions. 
This kind of procedure was realized
recently for MeOH~\cite{pusztai1}. Unfortunately, the experimental results
for EtOH and PrOH at high pressures are not available at present.
One can rely on the relevant data at 1bar~\cite{mendez-prop,mendez-prop2}
to get preliminary insights. At high pressures one may encounter
problems due to the rigidity of the models for EtOH and PrOH.
This issue requires additional considerations.
Nevertheless,
the coordinates files from simulations can be used to perform
analyses of the network of hydrogen bonds focusing on
the cooperative aspects of bonding in close similarity to \cite{pusztai1}.
Afterwards, necessary improvements of modelling at the level of united atom
and all atom type force fields would become plausible. 
We do not share the rather pessimistic opinion about the
present state of art of modelling of alcohols expressed in \cite{fomin} 
(obtained solely from the OPLS and COMPASS force fields for methanol under
high pressures)
and hope to make progress in solving some of the problems 
outlined above in future studies.

\section*{Acknowledgements}  
 Valuable discussions with Edgar Nu\~nez Rojas and
Imre Bak\'o  are gratefully acknowledged.

\newpage
\ukrainianpart

\title[Одноводневі спирти під тиском]
{Вплив тиску на властивості простих одноводневих спиртів. Уроки молекулярно-динамічного моделювання моделі UAM-EW типу об'єднаного атома}

\author[M. Агілар, Л. Пустай, O. Пізіо]
{M. Aгілар\refaddr{label1},
	Л. Пустай\refaddr{label2,label3},
	O. Пізіо\refaddr{label1}
}

\addresses{
	\addr{label1}
	Інститут хімії, Національний автономний університет Мексики, Сіркуіто Екстеріор, 04510, Мексика
	\addr{label2} Науково-дослідний центр фізики ім. Вігнера, H-1121, Будапешт,
	Конколі Теге Мут. 29--33, Угорщина 
	\addr{label3} Факультет передових наук і технологій, Університет Кумамото,
	2-39-1 Курокамі, Чуо-Ку, Кумамото, 860--8555, Японія
}

\makeukrtitle

\begin{abstract}
	\tolerance=3000%
		З використанням ізобарно-ізотермічного комп'ютерного моделювання молекулярної динаміки досліджується залежність від тиску низки властивостей простих одноводневих спиртів, а саме: метанолу, етанолу та 1-пропанолу.	Для кожного зі спиртів застосовується нещодавно запропоноване неполяризовне силове поле об'єднаного атома [\,V. Garc\'{i}a-Melgarejo et al., J. Mol. Liq., \textbf{323}, 114576 (2021)]. Точність силового поля оцінюється шляхом порівняння результатів моделювання та доступних з літератури експериментальних даних. Зокрема, досліджуються густина спиртів при збільшенні тиску, ізотермічна стисливість, статична діелектрична проникність та коефіцієнт самодифузії, починаючи від 1 бар до 3~кбар.	
	Обговорюється еволюція мікроскопічної структури під тиском з точки зору парних функцій розподілу та деяких координаційних чисел. У коментарях наведено висновки цього моделювання та необхідні зміни, які слід враховувати в майбутній роботі.

	\keywords молекулярна динаміка, метанол, етанол, 1-пропанол, тиск, густина, діелектрична проникність
\end{abstract}

\end{document}